\documentclass[12pt]{article}
\usepackage{graphicx}
\usepackage{amssymb}

\newcommand{\intk}{\int\frac{d^4 k}{(2\pi)^4}}

\newcommand{\beq}{\begin{equation}}
\newcommand{\eeq}{\end{equation}}

\newcommand{\q}{\mbox{${\mathbf q}$}}
\newcommand{\br}{\mbox{${\mathbf r}$}}

\newcommand{\ka}{\varkappa}
\newcommand{\la}{\lambda}
\newcommand{\de}{\delta}

\begin{document}


\begin{titlepage}

\begin{center}

{\bf \large Quantum power correction to the Newton law}

\vspace{1cm}

G.G. Kirilin\footnote{g\_kirilin@mail.ru} and I.B.
Khriplovich\footnote{khriplovich@inp.nsk.su}

\vspace{1cm}

Budker Institute of Nuclear Physics\\
630090 Novosibirsk, Russia\\
and Novosibirsk University

\end{center}

\bigskip

\begin{abstract}

We have found the graviton contribution to the one-loop quantum
correction to the Newton law. This correction results in
interaction decreasing with distance as $1/r^3$ and is dominated
numerically by the graviton contribution. The previous
calculations of this contribution to the discussed effect are
demonstrated to be incorrect.

\bigskip


\end{abstract}

\vspace{8cm}

\end{titlepage}

\section{Introduction}

The problem of corrections to the equations of motion, arising in
general relativity, is far from being new. The classical
relativistic corrections to these equations were found long ago by
Einstein, Infeld and Hoffmann \cite{eih}, and by Eddington and
Clark \cite{ec}. (A relatively simple derivation of these
corrections is presented in the textbook \cite{ll}.) Later this
result was reproduced by Iwasaki by means of Feynman diagrams
\cite{iw}. Thus, the problem of the classical relativistic
corrections to the Newton law is solved finally \footnote{Still,
erroneous papers on the subject are being published up to now. We
mean the articles \cite{kaz}, where the claim is made that the
classical relativistic corrections to the equations of motion of
two bodies separated by large distance depend essentially on the
inner structure of these bodies. We believe that this claim does
not withstand criticisms.}.

Let us note that the general structure of the relativistic classical correction to
the interaction potential of two bodies with masses $m_1$ and $m_2$, which would be
of second order in the Newton gravitational constant $k$, is clear immediately. Indeed,
the quantity $km/c^2$
($c$ is the velocity of light) has the dimension of length, so that with
the account for the symmetry under the interchange $m_1 \leftrightarrow
m_2$ the correction should be of the form
\beq
U_{cl}= a_{cl}\,\frac{k^2 m_1 m_2 (m_1 + m_2)}{c^2 r^2}.
\eeq
The dimensionless constant $a_{cl}$ as found in the above works equals $1/2$.

There is one more linear in $k$ combination of constants which can be used for the
construction of a power correction to the Newton potential. We mean
\[
\frac{k \hbar}{c^3}=l_p^2 \, ,
\]
where $\hbar$ is the Planck constant, $l_p=1.6\cdot 10^{-33}$ cm is the Planck length.
Clearly, such a correction, being of course of quantum nature, should look as follows:
\beq\label{vq}
U_{qu}= a_{qu}\,\frac{k^2 \hbar\, m_1 m_2}{c^3 r^3}.
\eeq
One has to find the numerical constant $a_{qu}$. In spite of extreme smallness of
the quantum correction, its investigation certainly has a methodological interest:
this is a closed calculation of a higher order effect in the nonrenormalizable
quantum gravity.

The reason why this problem allows for a closed solution is as
follows. The Fourier-transform of $1/r^3$ is
\beq\label{fur}
\int d\br \,\frac{\exp (-i \q \br)}{r^3}\,= -\,2\pi \ln q^2.
\eeq
This singularity in the momentum transfer $\q$ means that the
correction discussed can be generated only by diagrams with two
massless particles in the $t$-channel. The number of such
diagrams of second order in $k$ is finite, and their logarithmic
part in $q^2$ can be calculated unambiguously.

\begin{figure}
\begin{center}
\includegraphics*{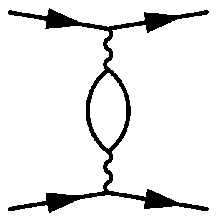}\hfill
\caption{Photon (neutrino) loop}\label{realloop}
\end{center}
\end{figure}

The corresponding diagrams with photons and massless neutrinos in
the loop (see Fig. \ref{realloop}) were calculated by Radkowski
\cite{rad}, Capper, Duff, and Halpern \cite{caph}, Capper and Duff
\cite{cap}, Duff and Liu \cite{dli}. This contribution to the
numerical factor $a_{qu}$ is
\beq\label{a1}
a_{\gamma \nu}=-\,\frac{4+N_\nu}{15\pi}\,,
\eeq
where $N_\nu$ is the number of massless two-component neutrinos.

As to the contribution to the effect from the graviton exchange,
it was considered by Donoghue [10--13], Muzinich and Vokos
\cite{muz}, Hamber and Liu \cite{ham}, Akhundov, Belucci and
ShiekhΠ\cite{akh}. However, there are no quantitative agreement
among the results of these works, even the predictions for the
sign of the correction differ.

We believe that the correct result for the quantum correction to
the Newton law will be sufficiently interesting from the
theoretical point of view. This is the aim of our investigation.
Among the previous works on the subject, the most detailed
presentation of the calculation is given in \cite{don1,akh}. Our
approach --- the direct calulation of Feynman diagrams, the choice
of the field operator for the gravitational field and of the gauge
--- is the same as in [10--13, 16]. It allows for a relatively detailed
comparison of calculations of separate contributions to the
effect. This comparison has demonstrated that in [10--13, 16] not
all diagrams are taken into account, and the considered
contributions are calculated incorrectly. Below, when discussing
concrete diagrams, we will come back to the comparison with the
previous works, including \cite{muz,ham}. And meanwhile, let us
note an obvious error in [10--13, 16]: therein formula for the
Fourier-transform of the function $1/r^3$ (see (\ref{fur}))
contains $\pi^2$, instead of $\pi$, and this error persists in the
final answer as well.

Some of the considered diagrams contribute also to the classical
relativistic correction. To check our calculations we computed in
parallel these classical contributions and compared them with the
corresponding results of \cite{iw}. As to these classical
corrections, we have complete agreement with \cite{iw} for each
diagram taken separately.

\section{Propagators and vertices}

We use below the units with $c=1$, $\hbar=1$.

As a field operator $h_{\mu\nu}$  we choose  the  deviation  of
the metrics $g _{\mu\nu}$ from the flat one:
\begin{eqnarray}
g_{\mu\nu}=\delta_{\mu\nu}+\ka\, h_{\mu\nu}\,; \qquad
\delta_{\mu\nu}={\rm diag}(1, -1,-1,-1); \qquad \ka^2=32\pi k\,=
32\pi l^2_p\,. \label{l:1}
\end{eqnarray}
We use the gauge where the graviton propagator is
\beq
D_{\mu\nu,\alpha\beta}(q)=i\,\frac{P_{\mu\nu,\alpha\beta}}{q^2+i0}\,;\qquad
P_{\mu\nu,\alpha\beta}\,=\,\frac{1}{2}\,(\delta_{\mu\alpha}\delta_{\nu\beta}
+\delta_{\nu\alpha}\delta_{\mu\beta}-
\delta_{\mu\nu}\delta_{\alpha\beta})\,.
\eeq
The tensor $P_{\mu\nu,\alpha\beta}$ is conveniently presented as \cite{hoo}
\[
P_{\mu\nu,\alpha\beta}\,=\,I_{\mu\nu,\alpha\beta} -
\,\frac{1}{2}\,\delta_{\mu\nu}\delta_{\alpha\beta}\,,
\]
where
$I_{\mu\nu,\alpha\beta}=\,\frac{1}{2}\,(\delta_{\mu\alpha}\delta_{\nu\beta}
+\delta_{\nu\alpha}\delta_{\mu\beta})$ is a sort of a unit operator with the property
\[
I_{\mu\nu,\alpha\beta}t_{\alpha\beta}=t_{\mu\nu}
\]
for any symmetric tensor $t_{\alpha\beta}$. Let us note the
following useful identity:
\beq\label{id}
P_{\alpha\beta,\ka\lambda} P_{\ka\lambda,\gamma\delta} =
I_{\alpha\beta,\gamma\delta}\,.
\eeq

Propagator of a scalar particle is the usual one:
\beq
G(p)=i\frac{1}{p^2+i0}\,.
\eeq

\begin{figure}
\begin{center}
\includegraphics*[width=0.90\textwidth]{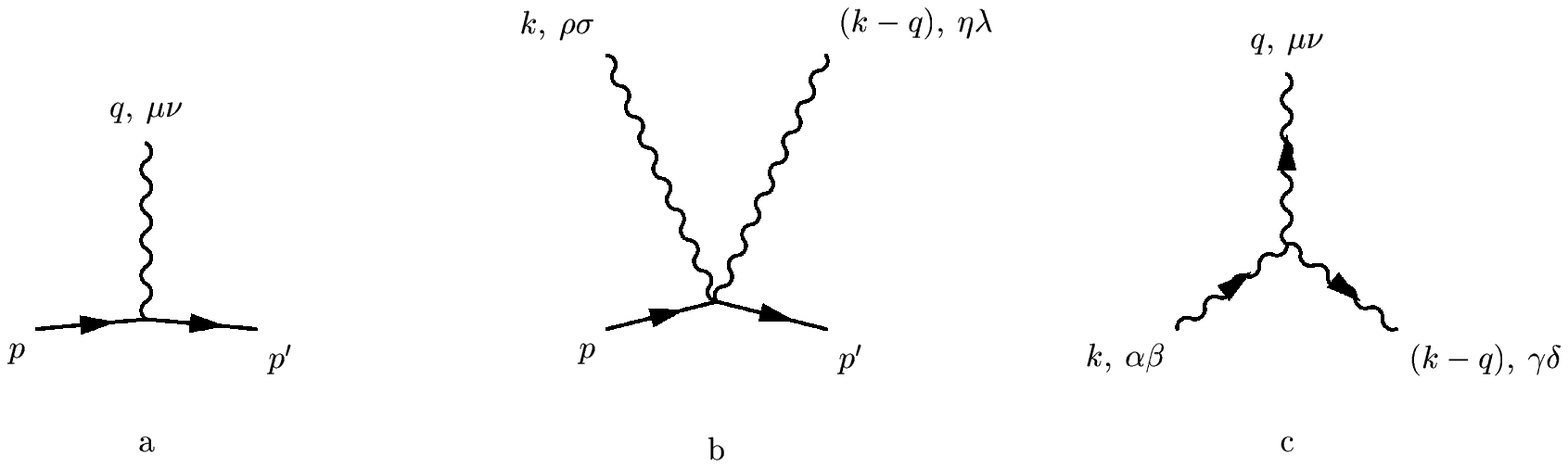}\hfill
\caption{Vertices}\label{vertex}
\end{center}
\end{figure}

Vertex of the interaction of a scalar particle with graviton is (see Fig.
\ref{vertex}a)
\beq\label{sc}
V_{\alpha\beta}(p,p')=-i \,\frac{\ka}{2}\left[p_\alpha {p'}_\beta
+ {p'}_\alpha p_\beta - \delta_{\alpha\beta}(pp'-m^2)\right].\nonumber \eeq

The contact interaction of a scalar particle with two gravitons (see Fig.
\ref{vertex}b) is
\begin{eqnarray}\label{v2}
&V_{\ka\lambda,\rho\sigma} = & i\ka^2\;
[\;I_{\ka\lambda,\alpha\delta}I_{\delta\beta,\rho\sigma}
(p_{\alpha}p'_{\beta}+p'_{\alpha}p_{\beta})-\,\frac{1}{2}\,
(\delta_{\ka\lambda}I_{\rho\sigma,\alpha\beta}+\delta_{\rho\sigma}
I_{\ka\lambda,\alpha\beta}) p_\alpha p'_\beta \nonumber \\
&&+\,\frac{(p'-p)^2}{4}\,(I_{\ka\lambda,\rho\sigma}
-\frac{1}{2}\delta_{\ka\lambda}\delta_{\rho\sigma})\;].
\end{eqnarray}
To our accuracy, one can neglect in this expression the last term,
with $(p'-p)^2$.

Let us note that in the works \cite{don1,akh} the vertex
(\ref{v2}) is erroneously presented (and indeed used in the
calculations) with a factor two times smaller, $\ka^2/2$ instead
of $\ka^2$. We will come back to this factor in Section 4.

The following useful identities are worth mentioning here:
\begin{eqnarray}
&&P_{\mu\nu,\alpha\beta}\left[p_\alpha {p'}_\beta + {p'}_\alpha
p_\beta - \delta_{\alpha\beta}(pp'-m^2)\right]=p_\alpha
{p'}_\beta + {p'}_\alpha p_\beta -\delta_{\alpha\beta}m^2\,,\label{pv1}\\
&& P_{\alpha\beta,\ka\lambda} P_{\gamma\delta,\rho\sigma}
V_{\ka\lambda,\rho\sigma}=
V_{\alpha\beta,\gamma\delta}\,.\label{pv2}
\end{eqnarray}

As to the 3-graviton vertex (see Fig. \ref{vertex}c), which has the most complicated
form, we will follow \cite{don1} and present it as
\begin{eqnarray}\label{v3}
&V_{\mu\nu,\alpha\beta,\gamma\delta}=&-i\,\frac{\ka}{2}\,\sum_i\,
{}^iv_{\mu\nu,\alpha\beta,\gamma\delta};\\ &
{}^1v_{\mu\nu,\alpha\beta,\gamma\delta}=&
P_{\alpha\beta,\gamma\delta}\,[k_\mu k_\nu+(k-q)_\mu
(k-q)_\nu+q_\mu q_\nu-
\frac{3}{2}\,\delta_{\mu\nu}q^2],\nonumber\\ &
{}^2v_{\mu\nu,\alpha\beta,\gamma\delta}=&2 q_\lambda q_\sigma [
I_{\lambda\sigma,\alpha\beta}I_{\mu\nu,\gamma\delta}
+I_{\lambda\sigma,\gamma\delta}I_{\mu\nu,\alpha\beta}-
I_{\lambda\mu,\alpha\beta}I_{\sigma\nu,\gamma\delta}
-I_{\lambda\nu,\alpha\beta}I_{\sigma\mu,\gamma\delta}],\nonumber\\
& {}^3v_{\mu\nu,\alpha\beta,\gamma\delta}=& q_{\lambda} q_{\mu}
(\delta_{\alpha\beta}I_{\lambda\nu,\gamma\delta}+
\delta_{\gamma\delta}I_{\lambda\nu,\alpha\beta})+ q_{\lambda}
q_{\nu} (\delta_{\alpha\beta}I_{\lambda\mu,\gamma\delta}+
\delta_{\gamma\delta}I_{\lambda\mu,\alpha\beta})\nonumber\\
&&-q^2(\delta_{\alpha\beta}I_{\mu\nu,\gamma\delta}+
\delta_{\gamma\delta}I_{\mu\nu,\alpha\beta})-
\delta_{\mu\nu}q_\lambda q_\sigma
(\delta_{\alpha\beta}I_{\gamma\delta,\lambda\sigma}+
\delta_{\gamma\delta}I_{\alpha\beta,\lambda\sigma}),\nonumber\\ &
{}^4v_{\mu\nu,\alpha\beta,\gamma\delta}=&
2q_\lambda[I_{\sigma\nu,\alpha\beta}I_{\gamma\delta,\lambda\sigma}(k-q)_\mu
+I_{\sigma\mu,\alpha\beta}I_{\gamma\delta,\lambda\sigma}(k-q)_\nu
\nonumber\\
&&-I_{\sigma\nu,\gamma\delta}I_{\alpha\beta,\lambda\sigma} k_\mu
-I_{\sigma\mu,\gamma\delta}I_{\alpha\beta,\lambda\sigma} k_\nu]
\nonumber\\ &&+
q^2(I_{\sigma\mu,\alpha\beta}I_{\gamma\delta,\sigma\nu}
+I_{\sigma\nu,\alpha\beta}I_{\gamma\delta,\sigma\mu})+
\delta_{\mu\nu}q_\lambda
q_\sigma(I_{\alpha\beta,\lambda\rho}I_{\rho\sigma,\gamma\delta}
+I_{\gamma\delta,\lambda\rho}I_{\rho\sigma,\alpha\beta}),\nonumber\\
& {}^5v_{\mu\nu,\alpha\beta,\gamma\delta}=&[k^2+(k-q)^2]\left(
I_{\sigma\mu,\alpha\beta}
I_{\gamma\delta,\sigma\nu}-\frac{1}{2}\,\delta_{\mu\nu}
P_{\alpha\beta,\gamma\delta}\right) \nonumber\\ && - k^2
\delta_{\gamma\delta}
I_{\mu\nu,\alpha\beta}-(k-q)^2\delta_{\alpha\beta}I_{\mu\nu,\gamma\delta}.\nonumber
\end{eqnarray}
In this vertex as well one can neglect, to our accuracy, the last
structure ${}^5v_{\mu\nu,\alpha\beta,\gamma\delta}$.

Let us note that in papers \cite{don1,akh} vertex (\ref{v3}) is
written down with an opposite sign. Our sign is confirmed by the
following argument: for physical gravitons with momenta $k$ and
$k-q$ vertex (\ref{v3}) should agree in the limit $q\rightarrow 0$
with the interaction (\ref{sc}) of a graviton with a scalar
particle.

\section{Simple loops}

\begin{figure}
\begin{center}
\includegraphics*{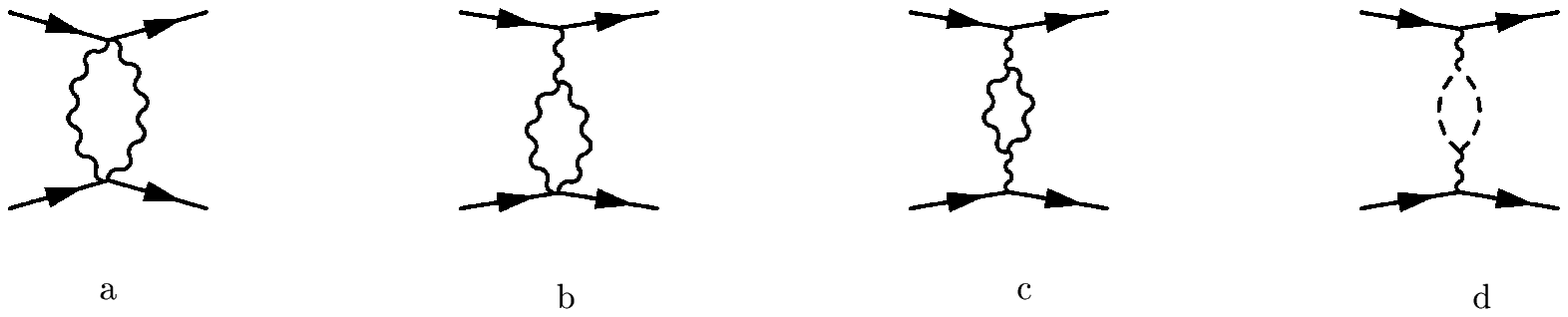}\hfill
\caption{Simple loops}\label{loop}
\end{center}
\end{figure}

It is convenient to start with the diagrams where the Feynman integrals contain
two denominators only.

The simplest of them, Fig. \ref{loop}a, is missed at all in
[10--13, 16]. Its calculation causes no difficulty, one has only
to use the identity (\ref{pv2}) and to go over to the
nonrelativistic limit in both 2-graviton vertices. The result for
this contribution to the quantum correction is
\beq\label{qu1}
U_{qu1}=-\,\frac{22}{\pi}\,\frac{k^2 m_1 m_2}{r^3}\,.
\eeq

The calculation of the next diagram, Fig. \ref{loop}b, and that
obtained from it by interchanging scalar particles, is also
sufficiently simple and results in
\beq\label{qu2}
U_{qu2}=\,\frac{26}{3\pi}\,\frac{k^2 m_1 m_2}{r^3}\,.
\eeq
The result of \cite{don1} for this contribution differs from
(\ref{qu2}) only by a wrong power of $\pi$. The corresponding
result of \cite{akh} is quite different.

As to the diagrams Fig. \ref{loop}c,d with the polarization
operator of graviton, we do not have much to add to works [10--13]
on the method of the calculation, and nothing to add at all on
the result itself (which is used also in \cite{muz, akh}).
However, for the completeness we present here briefly this
calculation.

The effective Lagrangian corresponding to the sum of these
diagrams, with gravitons and vector ghosts, as obtained by 't
Hooft and Veltman \cite{hoo}, is
\beq
L =-\,\frac{1}{16\pi^2} \ln |\,q^2| \left(\frac{1}{120}\,R^2 +
\frac{7}{20}\,R_{\mu\nu}R^{\mu\nu}\right).
\eeq
To linear approximation the Ricci tensor and scalar curvature of the external
gravitational field which enter this expression are
\[
R_{\mu\nu}=\,\frac{\ka}{2}\,h_{\alpha\beta}(q^2
I_{\mu\nu,\alpha\beta}+q_{\mu} q_{\nu} \delta_{\alpha\beta} -
q_{\mu} q_{\alpha} \delta_{\nu\beta} - q_{\nu} q_{\alpha}
\delta_{\mu\beta})=\,\frac{\ka}{2}\,h_{\alpha\beta}
r_{\mu\nu,\alpha\beta}\,,
\]
\[
R=\ka h_{\alpha\beta}(q^2 \delta_{\alpha\beta} - q_{\alpha}
q_{\beta}) = \ka h_{\alpha\beta} r_{\alpha\beta}\,.
\]
The corresponding contribution to the graviton polarization operator is
\beq
\Pi_{\alpha\beta,\gamma\delta}=-\,\frac{\ka^2}{8\pi^2} \,\ln
|\,q^2| \left(\frac{1}{120}\,r_{\alpha\beta}r_{\gamma\delta} +
\frac{7}{80}\,r_{\mu\nu,\alpha\beta}r_{\mu\nu,\gamma\delta}\right).
\eeq
We have taken into account here two possibilities of identifying
$R_{\mu\nu}$ and $R$ with the upper and lower external gravitons.
The subsequent calculation is straightforward. Let us mention only
that the summation over $\mu$, $\nu$ is conveniently performed at
the end. Finally, this contribution to the quantum correction is
\beq\label{qu3}
U_{qu3}= -\,\frac{43}{30\pi}\,\frac{k^2 m_1 m_2}{r^3}\,.
\eeq

Let us mention that diagrams \ref{loop}c,d were computed in other
variables, $\psi_{\mu\nu}=h_{\mu\nu}-1/2\de_{\mu\nu}h$, in Refs.
\cite{ham,capl,duf}, and in the source description of gravity due
to Schwinger in \cite{rad}.

\section{Triangle diagrams}

\begin{figure}
\begin{center}
\includegraphics*{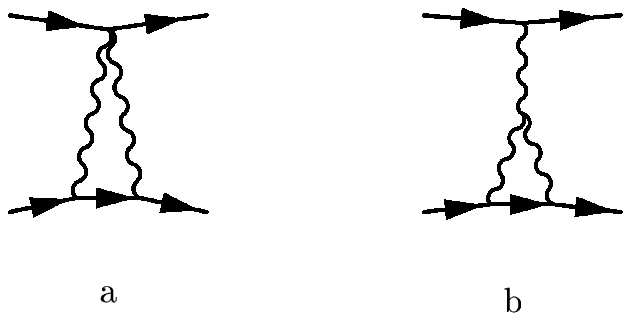}\hfill
\caption{Triangle diagrams}\label{triangle}
\end{center}
\end{figure}

A sort of master formula for triangle diagrams, Figs.
\ref{triangle}a,b, reads (we keep here only terms singular in
$|\,q|$):
\beq\label{mft}
i\intk \;\frac{1}{k^2 (k-q)^2
\left((p-k)^2-m^2\right)}=\frac{1}{32 \pi^2 m^2}
\left(\,\frac{\pi^2 m}{ \sqrt{|\,q^2\,|}}+\ln|\,q^2\,| \right).
\eeq
It is conveniently obtained by calculating first the imaginary
part of its lhs in the $t$-channel, and then restoring its rhs via
the dispersion relation. The first term in the rhs of formula
(\ref{mft}) generates $1/r^2$ in the coordinate representation
and contributes to the classical relativistic correction. It is
kept in (\ref{mft}) to check the calculations by comparison with
the corresponding results of \cite{iw}.

Our result for the contribution of more simple diagrams of the
type Fig. \ref{triangle}a is
\beq\label{qu4}
U_{qu4}=\,\frac{28}{\pi}\,\frac{k^2 m_1 m_2}{r^3}\,.
\eeq
This contribution is also missed in \cite{don1,akh}.

These diagrams contribute to the classical correction as well. An
extra proof of our normalization for the seagull vertex is the
agreement with the corresponding classical result of~\cite{iw}.

Much more tedious is the calculation of diagrams of the type Fig.
\ref{triangle}b. It results in
\beq\label{qu5}
U_{qu5}=-\,\frac{29}{3\pi}\,\frac{k^2 m_1 m_2}{r^3}\;.
\eeq
The corresponding result of \cite{akh} differs from (\ref{qu5})
only by sign and wrong power of $\pi$. The result of \cite{don1}
for this contribution is quite different.

\section{Box diagrams}

\begin{figure}
\begin{center}
\includegraphics*{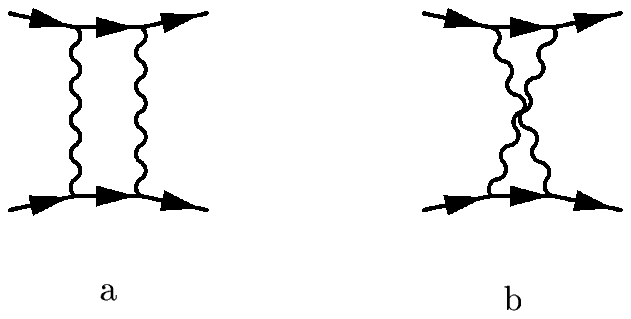}\hfill
\caption{Box diagrams}\label{box}
\end{center}
\end{figure}

The expressions for the matrix elements corresponding to the box
diagrams Figs. \ref{box}a,b can be presented as
\begin{eqnarray}\label{s}
M_{s}=i\;\frac{\ka^4}{16 m_1 m_2}\intk\;\frac{[
a-b(D_{1}+D_{2})]^2 }{k^2 (k-q)^2 D_{1} D_{2}}\;;
\end{eqnarray}
\[
D_{1}=k^2-2(p_1 k),\quad D_{2}=k^2+2 p_2k,\quad a=2(p_1 p_2)^2-m^4,\quad b=(p_1
p_2).
\]
\begin{eqnarray}\label{u}
M_{u}=i\;\frac{\ka^4}{16 m_1 m_2}\intk\;\frac{[
a'+b'(D_{1}+D'_{2})]^2 }{k^2 (k-q)^2 D_{1} D'_{2}}\;;
\end{eqnarray}
\[
D'_2=k^2-2 (p_2+q, k), \quad b'=(p_1, p_2+q), \quad a'=2(p_1, p_2+q)^2-m^4.
\]
It is convenient to single out in the numerators of these
integrals such structures that cancel one or both denominators
$D_{1},\; D_{2}\;(D'_{2})$. When cancelling a single denominator,
one is left with effectively triangle diagrams of the type Fig.
\ref{triangle}a:
\beq\label{s1}
M_{1s}=i\;\frac{\ka^4}{16 m_1 m_2} \int \frac{d^4 k}{(2\pi)^4 k^2
(k-q)^2} \left\{b^2\left(\frac{2(p_2 k)}{D_{1}}-\frac{2(p_1
k)}{D_{2}}\right) -2 a b
\left(\frac{1}{D_{1}}+\frac{1}{D_{2}}\right)\right\};
\eeq
\beq\label{u1}
M_{1u}=i\;\frac{\ka^4}{16 m_1 m_2}\int \frac{d^4 k}{(2\pi)^4 k^2
(k-q)^2}\left\{ {b'}^2\left(-\frac{2(p'_2 k)}{D_1}-\frac{2(p_1
k)}{D'_2}\right) +2
a'b'\left(\frac{1}{D_1}+\frac{1}{D'_2}\right)\right\}.
\eeq
It can be easily demonstrated that to the accuracy we are
interested in, expressions (\ref{s1}) and (\ref{u1}) cancel.

Expressions with two denominators deleted,
\beq\label{s2}
M_{2s}= i\;\frac{\ka^4}{16 m_1 m_2} \intk \;\frac{2 b^2}{k^2
(k-q)^2}\;,
\eeq
\beq\label{u2}
M_{2u}= i\;\frac{\ka^4}{16 m_1 m_2}\intk \;\frac{2 {b'}^2}{k^2
(k-q)^2}\;,
\eeq
correspond to the diagrams of the type Fig. \ref{loop}a. These
contributions add up into the
following result for the effect discussed:
\beq\label{qu6}
U_{qu6}=-\,\frac{8}{\pi}\,\frac{k^2 m_1 m_2}{r^3}\,.
\eeq

Now we are left with the ``irreducible'' parts of diagrams Figs.
\ref{box}a,b. These irreducible matrix elements are conveniently
obtained by calculating first their imaginary parts, in the $s$
and $u$ channels respectively, and then restoring the real parts
through the dispersion relations. The results are (we omit
nonsingular in $|\,q^2\,|$ terms)
\begin{eqnarray}\label{s3}
M_{0s}&=&i\;\frac{\ka^4 a^2}{16 m_1 m_2}\intk\;
\frac{1}{(k^2-\la^2) ((k-q)^2-\la^2) (k^2-2p_1k)(k^2+2p_2k)}
\\&=& -\,\frac{\ka^4 a^2}{(16 m_1
m_2)^2|\,q^2\,|\pi^2}\left[-1+\frac{s-2 (m_1^2+ m_2^2)}{6 m_1
m_2}\right] \:\ln\frac{|\,q^2\,|}{\la^2}\;;
\end{eqnarray}
\begin{eqnarray}\label{u3}
M_{0u}&=&i\;\frac{\ka^4 a^2}{16 m_1 m_2} \intk \;
\frac{1}{(k^2-\la^2) ((k-q)^2-\la^2)
(k^2-2p_1k)(k^2-2(p_2+q,k)}\\&=& -\,\frac{\ka^4 a^2}{(16 m_1
m_2)^2|\,q^2\,|\pi^2}\left[1+\frac{u}{6 m_1 m_2}\right]
\:\ln\frac{|\,q^2\,|}{\la^2}\;.
\end{eqnarray}
In the above formulae $s=(p_1+p_2)^2$ and $u=(p_1-p_2-q)^2$.
Expressions (\ref{s3}), (\ref{u3}) are convergent in the
ultraviolet sense, but diverge in the infrared limit, depending
logarithmically on the ``graviton mass'' $\lambda$. As usual, such
behaviour is directly related to the necessity to cancel the
infrared divergence in the Bremsstrahlung diagrams (of course, the
gravitational Bremsstrahlung in the present case). Though the
leading singularity in $q$ is of the type $\ln |\,q^2|/|\,q^2|$, a
term with $\ln |\,q^2|$ arises in the sum of the irreducible boxes
as well, and generates the following quantum correction to the
Newton potential:
\beq\label{qu7}
U_{qu7}=-\,\frac{23}{3\pi}\,\frac{k^2 m_1 m_2}{r^3}\,.
\eeq
It is worth mentioning that, as distinct from the previous
contributions where $|\,q^2|$ served as an infrared cut-off for
ultraviolet divergent integrals, here $|\,q^2|$ is the upper limit
for infrared divergent integrals.

For the box diagrams as well, we have checked that our results for
thus generated classical corrections agree completely with those
of \cite{iw}.

The box contributions to the quantum correction are missed at all
in [10--13, 16], though diagrams Fig. \ref{box}a,b are considered
in \cite{dot} from another point of view.

On the other hand, neither in \cite{muz}, nor in \cite{ham} we
could find any mention of the ``infrared'' contribution of the
type (\ref{qu7}). In fact, in \cite{ham} the problem of classical
and quantum corrections was treated in different variables,
$\psi_{\mu\nu}=h_{\mu\nu}-1/2\de_{\mu\nu}h$. It can be easily
demonstrated that the expressions for the box diagrams are exactly
the same in both variables, $\psi$ and $h$. However the box
contributions, as calculated in \cite{ham}, disagree both with the
classical ones obtained in \cite{iw} (which are demonstrated
explicitly in \cite{iw} to be the same in both variables, $\psi$
and $h$) and with our results for the quantum correction, be it
(\ref{qu6}), or (\ref{qu7}), or the sum of (\ref{qu6}) and
(\ref{qu7}).

At last few words more on Ref. \cite{muz}. The approach advocated
therein looks quite interesting and promising. However, the
results for the quantum correction presented in \cite{muz} do not
agree with ours (neither do they agree with those of [10--13, 15,
16]). Due to the lack of details in \cite{muz}, we cannot say with
certainty what is the origin of the disagreement. Still, an
impression arises that at least it is overlooked in \cite{muz}
that the irreducible triangle diagrams generate not only classical
corrections, but quantum corrections as well, i.e. it seems that
in \cite{muz} the second term is omitted in formula (\ref{mft}).

\section{Conclusions}

Summing up all the contributions obtained, (\ref{qu1}),
(\ref{qu2}), (\ref{qu3}), (\ref{qu4}), (\ref{qu5}), (\ref{qu6}),
(\ref{qu7}), we arrive at the following result for the Newton
potential, with the discussed quantum correction due to the
two-graviton exchange included:
\beq\label{f}
U(r)=-\,\frac{k m_1 m_2}{r}\,\left(1+\,\frac{121}{10\pi}\, \frac{k
\hbar}{c^3 r^2}\right).
\eeq
Let us note that the derived overall correction enhances, but not
suppresses the common Newton attraction.

\section*{Acknowledgements}
We are grateful to S. Deser and M.J. Duff for bringing to our
attention Refs. \cite{rad,dli,duf} and for interesting comments.
The investigation was supported in part by the Russian Foundation
for Basic Research through Grant No. 01-02-16898, through Grant
No. 00-15-96811 for Leading Scientific Schools, by the Ministry of
Education Grant No. E00-3.3-148, and by the Federal Program
Integration-2002.

\end{document}